\documentstyle[12pt,psfig,epsf,alltt]{article}

\begin{document}
\begin{flushright}
BI-TP 2003/17\\
SFB/CPP-03-27
\end{flushright}

\begin{center}                                                                  
{\Large \bf 
Parallel Computation of Feynman diagrams with DIANA}
\end{center}

\vspace{4mm}
\begin{center}
M.~Tentyukov$^{a~b~\dag~\ddag}$
and
J.~Fleischer$^{b}$
\end{center}

\begin{flushleft}\em
~~~~~~~~~~~a) Institut f\"ur Theoretische Teilchenphysik,  Universit\"at
Karlsruhe,  Germany\\[1mm]
~~~~~~~~~~~b)Fakult\"at f\"ur Physik, Universit\"at Bielefeld, Germany\\[5mm]
~~~~~~~~~~~\dag) On leave from
Joint Institute for Nuclear Research, Dubna,
Russia\\[1mm]
~~~~~~~~~~~\ddag) Supported by
DFG under FL 241/4-2
\end{flushleft}

\begin{abstract}
Co-operation of the Feynman DIagram ANAlyzer (DIANA) with the underlying
operational system (UNIX) is presented. We discuss operators to run
external commands and a recent development of parallel processing
facilities and an extension in the spirit of a component model.
%\vspace{1pc}
\end{abstract}
%\end{frontmatter}

\section{Introduction}

We have developed the program DIANA (DIagram ANAlyser),
which allows to calculate Feynman diagrams as `input'
for further formulae manipulating languages (mainly FORM
\cite{FORM}) \cite{Avdeev:1996ad}, \cite{Tentyukov:1999is}
(see also http://www.physik.uni-bielefeld.de/\verb|~|tentukov/diana.html).
Meanwhile there exists a series of applications \cite{Anwend},
which would not have been possible without DIANA.

The problem is that in order to cope with the high precision experiments
of present-day high energy accelerators, the corresponding
calculations within the `Standard Model' (SM) of elementary particle physics
must be of similar precision. This means, e.g., for 1-loop approximations
even for $2 \to 2$ processes the calculation of
the order of several hundred diagrams. This number increases
rapidly with the number of external legs and/or the number of loops
such that one is easily forced to calculate a number of diagrams of
the order of thousands. For this one needs an automation to generate
the diagrams and this is what DIANA does.

Having produced the FORM `input' for each of the diagrams, in a next step
the execution of the FORM program has to run for each diagram separately
and the analytic result of all the separate calculations has to be summed up.
In many cases like e.g. 1-loop $2 \to 2$ calculations the execution time
of the FORM jobs is still relatively short and one can just run the
corresponding
jobs one after the other (e.g. in terms of `folders'). In more complicated
cases, however, not only the number
of diagrams increases but also the analytic calculation of the single
diagrams gets more and more time consuming. Here it pays to distribute the
essentially independent calculations of the separate diagrams over different
computers. In the project described here we have extended DIANA to be
able to perform this task and we give detailed hints for potential users
of how to apply it.

The program DIANA, written in C, contains two ingredients:
\begin{enumerate}
\item
Analyzer of diagrams.
\item
Interpreter of a special text manipulating (TM) language.
\end{enumerate}
The analyzer reads QGRAF \cite{QGRAF} output and passes the necessary
information to the interpreter. For each diagram the interpreter performs
the TM program, producing input for further evaluation (e.g. by means of FORM)
of the diagram.
The TM language is a TeX-like language.
The main goal of the language is the creation of text files.

Similar to the TeX language, each command has to begin with an escape
(``$\backslash$'') character. All lines without escape - characters are
simply typed to the output file.
Some of the TM  commands are built-in TM operators
while some are user-defined functions (returning a value) written in
the TM language itself.

The main goal of parallel processing
is to reduce wall-clock time\footnote{The elapsed time between
start and finish of a process.} which is the users
waiting time.
Parallelism doesn't come for free; always it has some
overhead with respect to serial execution, but it can significantly reduce
the wall-clock time.

Not every problem can be divided into parallel tasks. An example of a
parallelizable problem is the multiplication of two matrices. An example of a
non-parallelizable problem is the calculation of the Fibonacci series
(1,1,2,3,5,{\ldots}) by means of the formula $F(k+2)=F(k+1)+F(k)$.

Computations in perturbative Quantum Field Theory involve the calculation of Feynman
diagrams. DIANA is used to produce the input for their further analytic evaluation.
In fact DIANA needs a ``model file'' produced by the user and then generates
the ``input'' in terms of formally defined vertices and propagators and of
coupling parameters \cite{Tentyukov:1999is}.
This generation of Feynman diagrams itself is not a good parallelizable problem. Moreover,
the time DIANA spends to produce such input is negligible with respect to
the time used for the
diagram evaluation. Thus the present approach consists in generating
input for each diagram sequentially, with the possibility to parallelize the further
evaluation.

After the diagrams are generated, DIANA may be used as a ``control
center'' for further evaluations (in parallel) by  invoking the
corresponding programs and providing them the generated input. To avoid a
cluster/processor overloading, each time only one job per node/processor
is actually running while all the rest is queued.
Such an
approach is known as a Batch Queueing System (BQS).
There exist a lot of such systems, both free and commercial \cite{NASAQueuing}.
Usually, these systems are
aimed to deal with a big number of different isolated tasks, while in our case
we have a huge number of nearly even tasks.

There are two kinds of optimization of BQS available. The first is for the case
when the average number of jobs is much larger than the number of nodes. In
this case BQS usually provides various complicated methods for scheduling,
message pasting, job run-time quotas and resource management in order to
distribute computational resources among various tasks of all kinds.
In our case all these mechanisms are not needed.
Secondly, if the average number of jobs is comparable with the number of
nodes, which is typical for parallel computation, a BQS should provide load
balancing, process migration mechanisms and some others.

Compared to that, DIANA has to implement
parallelism in terms of job queueing, and the typical situation consists of
a long queue of nearly even tasks with the same priority. Evidently, we
need not any load balancing. Indeed, the best procedure is to send a new task
to a node immediately when the node becomes free.
Thus, the real problem in our case is the problem of synchronization.
Traditionally, this problem is ignored in BQS since all jobs are assumed to
be completely independent. The synchronization problem arises in true
parallel systems. Since we use queueing in order to implement a parallelism,
the problem is of immediate interest in our case.

The TM language contains several groups of operators permitting DIANA to
make a fine co-operation with the underlying operational system (UNIX).
The operators \verb|\system()| and \verb|\asksystem()| execute an external
command synchronously (see sect. \ref{synchoperators}), i.e. they wait for
the command to be completed. The
operator \verb|\pipe()| (sect. \ref{thepipe}) starts the external
command opening an input-output
channel for it, and can be used in order to extend DIANA in the spirit of a
component model (sect. \ref{compmod}).
The term ``component model'' (see,
e.g., \cite{componentmod}) means
that a software system is built from ``components''.
Components are high level aggregations of smaller software
pieces, and provide a ``black box'' building
block approach to software construction.

Let us now suppose that we use a cluster with disk space shared by means of
NFS\footnote{Network File
System},  and all results must be
collected into one resulting file \verb|log.all|, while every job produces a
file \verb|log.#| where \verb|#| is the order number of the job. Since jobs
are running and completing independently of each other, we can only
collect  all \verb|log.#| into \verb|log.all| after {\em all} the jobs are
finished. This leads to producing a lot of files \verb|log.#| at
an intermediate step, which can overload a file system.

The simple solution is to allow some of the newly started jobs to be synchronized
with all previously started jobs. Indeed, in this case after each job we can
start another ``slave'' job, which appends the \verb|log.#| file containing the
result of the ``master'' job to the file \verb|log.all|. To obtain a correct
order of the file \verb|log.all|, this copying job must be performed only
after all earlier jobs are finished.

Another problem with cluster computations is that the optimal placement of
the resulting file \verb|log.#| is usually a \verb|/tmp| directory which is
local with regard to the
current node, on which the job ``number \verb|#|'' is performed.
But to do this, the ``slave'' copying job has to ``stick'' to the ``master'' job,
i.e.  it must be performed on the same node as the ``master'' job.

Both above mentioned problems are solved by means of so-called job
attributes, see sect.~\ref{attribs}.

Yet another problem is the transaction problem, which in our case means
whether a job group
should be restarted on fa\-i\-lure, and what the fa\-i\-lure/success conditions are.
Sometimes it is
necessary to perform some complicated job with many unreliable
intermediate steps (e.g., breakdown of some connection) and at the end
return the exit code depending on the status of the whole transaction. In
such cases it is useful to have a possibility to restart the job
many times. But for conventional jobs consisting of one external program the
number of restarts must not be too large in order to permit the user to kill
jobs which failed.

Conditions on which a job is assumed to have failed may be rather complicated,
too.
For example, the above mentioned copying ``slave'' job is just a quick system
command ``cp''. The probability of its failure is very low, less then the
probability of some network connection failure during the commands
initiation. If started, such a job, most probably, will be completed
successfully independently of the DIANA client/server status. That is why it
is reasonable to assume that the job is successful if it is started by  the
server. On the other hand, the ``master'' job is expected to return {\em some} exit
code, but not to be killed by some signal.
For the solution of these problems see sect.~\ref{attribs} as well.

The operator \verb|\_exec()| and companions can be used to define several
macros and TM functions in order to create a BQS which meets our requirements;
for a simplified example see sect.~\ref{parascript}. There is the possibility to
build a really powerful BQS based on the \verb|\_exec()| family combined with
external components communicating with DIANA through the operator
\verb|\pipe()|. A simplified version of this is demonstrated.

%%%%%%%%%%%%%%%%%%%%%%%%%%%%%%%%%%%%%%%%%%%%%%%%%%%%%%%%%%%%%%%%%%%%%%%%%%%%

\section{Running external commands}
\label{synchoperators}
The operator
\verb|\system(command)| executes the external \verb|command| and returns the exit
code. Usually, 0 means that the invoked program was executed successfully.
Example: \verb|\system(ls)| types to the screen the content of the current
directory and returns 0.

The operator
\verb|\asksystem(command,text)|
executes the external \verb|command|, using the second argument as its
``standard input''. It reads the first line of the external command output and
stores it. Then the external command is terminated, and the stored line is
returned. For example, \verb|\asksystem(cat,Hello!)| returns the text
``Hello!'', but \verb|\asksystem(cat,One!\eol()Two!)| returns the text
``One!'' (\verb|\eol()| is the DIANA command ending the line).

Another example:
\begin{verbatim}
\offleadingspaces
  * Host: \asksystem(hostname,); \asksystem(date,)
\end{verbatim}
in the TM program has led to the following line in the
produced output:
\begin{verbatim}
* Host: phya24; Fri May  30 22:51:38 CET 2003
\end{verbatim}

\section{Dialog with an external program}
\label{thepipe}
\subsection{Starting a pipe}
The operator \verb|\asksystem()| is used to ``ask'' a single question from
the command, as described above.
In order to organize a dialog with an external program, the operator
\verb|\pipe(command)| can be used. It starts the \verb|command|, swallowing its standard
input and output. It returns an integer number, a
PID\footnote{Process IDentifier} of a newly started process, or an empty
string, if it fails. Further, the returned PID can be used in order to send
some text to the program, or to read  program output.

After the operator \verb|\pipe()| returns the control to the TM program, the
program initiated by \verb|command| continues to run. Its standard input and output
references are
stored by DIANA. Both input and output are not
connected with any terminal device. The running
external program is visible from the TM program as a duplex ``pipe'' identified
by the  PID of the external program.

The operator \verb|\closepipe(rpipe)| (here \verb|rpipe| is the PID of the external
program returned by \verb|\pipe()|) terminates the external program. It
closes the programs' IO channels, sends to the program a KILL
signal
and waits for the \verb|command| to be
finished.

\subsection{Reading and writting with the pipe}
There are two operators to write to the pipe. One is
\verb|\topipe(rpipe,text)|, and another is \verb|\linetopipe(rpipe,text)|. The
latter appends a new line symbol to the text placed into the pipe while the
former sends \verb|text| to the pipe as it is.
The necessity of two operators comes from the fact that most of interactive
programs have line-oriented input/output. The program reads from the
terminal a whole line and will not continue until the line is completed.
On the other hand, sometimes it is necessary to use character-oriented 
input/output.

 The
corresponding reading operators  are
\verb|\frompipe(rpipe)| and  \linebreak \verb|\linefrompipe(rpipe)|.
The operator \verb|\linefrompipe()| reads from the pipe and returns a
whole line. The operator is fast, but it is aimed to read only
lines. In contrast, the operator \verb|\frompipe()| can be used in order
to read parts of lines. It stops reading if the line is completed,
at an end-of-file condition or if there are no more data
available in the pipe. The operator \verb|\frompipe()| reads from the pipe
character-by-character, therefore it is much slower then the operator
\verb|\linefrompipe()|\footnote{The operator $\backslash${\tt{}linefrompipe()}
makes bufferred
reading. It reads some block from the pipe, and then returns the part of the
block corresponding the first whole line. After the operator is performed,
some piece of the text corresponding to the beginning of the next
line
is also
transferred from the pipe into an internal buffer.}.

It is not allowed to mix \verb|\frompipe()| and \verb|\linefrompipe()|. The
result may be unpredictable and probably not what one expects.

These operators may be used to deal with the DIANA standard input/output.
If the first argument, \verb|rpipe|, is $\leq 0$, the operators
\verb|\topipe(rpipe,text)| and  \linebreak \verb|\linetopipe(rpipe,text)| will output
their second argument to the terminal. The same with \verb|\frompipe()|
and \verb|\linefrompipe()|: if their arguments are $\leq 0$, these
operators read from the keyboard.

\subsection{Related operators}
The operators \verb|\frompipe()| and \verb|\linefrompipe()| block the TM
program until some data will be available from the pipe. Sometimes this is not
acceptable. In order to solve this problem, the user can apply
the operator \verb|\ispipeready(rpipe)| (is-pipe-ready). The operator
returns \verb|ok| if there are
data available for the \verb|\frompipe()| operator. If
\verb|\ispipeready()| returns an empty string, the operator
\verb|\frompipe()| invoked afterwards will block the TM program 
and wait for some data to be available.

The operator \verb|\ispipeready()| may be used only in combination with
\linebreak \verb|\frompipe()|. The operator \verb|\linefrompipe()| may still return
some text immediately even if the operator \verb|\ispipeready()| has
returned an empty string.

The operator \verb|\checkpipe(rpipe,signum)| sends the signal ``\verb|signum|''
(an integer number) to the external
program initiated by \verb|\pipe()|. If the signal is delivered successfully,
the operator returns
\verb|ok|. If the signal cannot be delivered, e.g. if the external program was
terminated, then the operator returns an empty string.
If the first argument, \verb|rpipe|, is $\leq 0$, the operator does nothing
and returns \verb|ok| since DIANA assumes that the user checks the standard
input/output. If the second argument, \verb|signum|, is $< 1$,
it is assumed to be 0. It must be an
integer number, symbolic names are not accepted. The primary purpose of the
operator is to send the signal 0 to the pipe: this signal will not affect the
program  but all checking will be performed.

\subsection{Extension of DIANA by means of external components}
\label{compmod}
The operator \verb|\pipe()| can be used  to add some functionality
to DIANA in the spirit of a component model (see Introduction).

Let us suppose that we need to deal with high precision arithmetic.
Under UNIX, there exists the command ``\verb|dc|'' -- a standard 
desk calculator which supports
unlimited precision arithmetic.
It reads arguments from the standard input as so-called ``reverse-polish
notation''\footnote{Reverse Polish Notation is an arithmetic formula
notation, introduced in 1920 by the Polish mathematician Jan Lukasiewicz.
In this notation, the operands precede the operators, thus
dispensing the need of parentheses. For
example, the expression 3 * ( 4 + 7) would be written as 3~4~7~+~*. } and types
the answer to the standard output.

The following function \verb|\fdiv(a,b)| returns the quotient \verb|a|/\verb|b|
with 10 fraction digits:
\begin{verbatim}
\function fdiv a,b;\modesave()\-
   \if not exist"_dc"then
      \if"\export(_dc,\pipe(dc))"eq""then
         \killexp(_dc)
         \moderestore()\return()
      \endif
      \linetopipe(\import(_dc),\(10 k))
   \endif
   \let(a,\tr(-,_,\get(a)))
   \let(b,\tr(-,_,\get(b)))
   \linetopipe(\import(_dc),\(c )\get(a)\( )\get(b)\( / p))
   \moderestore()\return(\linefrompipe(\import(_dc)))
\end
\end{verbatim}
This is a very simple demo version, it is non-optimal and does not
contain exception handling.

First, if \verb|dc| is not running, the function starts \verb|dc|, swallowing its
input-output channels (it assumes that the external program \verb|dc| is
started if the global variable \verb|_dc| is defined).
Then it sets the precision 10, sending the line
``\verb|10 k|'' to \verb|dc|.

It replaces possible signs ``\verb|-|'' in both \verb|a| and \verb|b|
by the symbol ``\verb|_|'': the input \verb|dc| language requires negative numbers
to be preceded by ``\verb|_|''; the sign  ``\verb|-|'' cannot be used for this,
since it is a binary  operator  for  subtraction.

Then the function sends the line ``\verb+c |a| |b| / p+'' to the running \verb|dc|
command.
The calculator clears
previous results (command \verb|c|), evaluates the quotient and puts
the result to the standard output (command \verb|p|).

In this example it is possible to use the operator\\
\verb+\asksystem(dc, 10 k |a| |b| / p)+\\
instead: 
\begin{verbatim}
\function fdiv a,b;\modesave()\-
   \let(a,\tr(-,_,\get(a)))
   \let(b,\tr(-,_,\get(b)))
   \moderestore()\return(
     \asksystem(dc, \(10 k )\get(a)\( )\get(b)\( / p))
   )
\end
\end{verbatim}

For each call of \verb|\fdiv(a,b)| it
would start \verb|dc|, evaluate and return the result. The time consumed
by such a function will be about 100 times larger than the variant with
\verb|\pipe()|, but it is possible in principle.

Another example illustrates the ``real'' dialog with the component used
to organize some simple GUI\footnote{Graphical User Interface}.
Let's suppose some Tcl/Tk\footnote{A popular scripting language
intended primarily for creating  GUI, see http://www.tcl.tk/scripting}
script is placed into the
executable file \verb|guiDemo| (see appendix \ref{guiDemo}; this
example is available at\\
 http://www.physik.uni-bielefeld.de/\verb|~|component\_model.tar.gz).

\begin{figure}[ht]
\begin{minipage}[b]{.32\linewidth}
a) \centerline{\epsfxsize=35mm \epsfbox{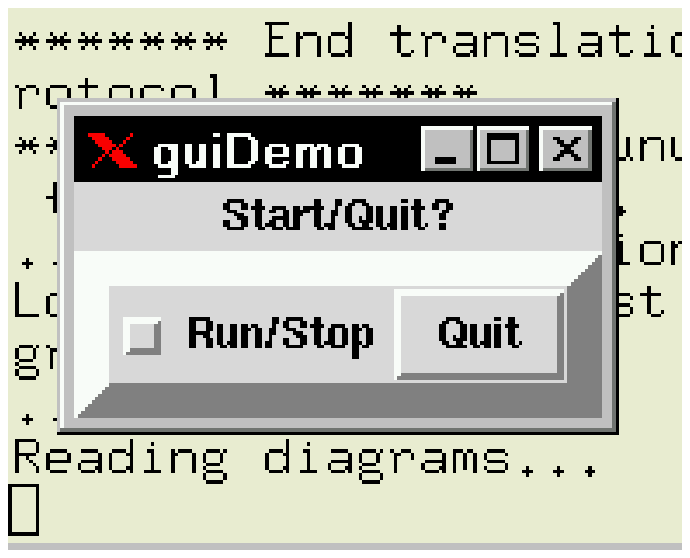}}
\end{minipage}
\begin{minipage}[b]{.32\linewidth}
b) \centerline{\epsfxsize=35mm \epsfbox{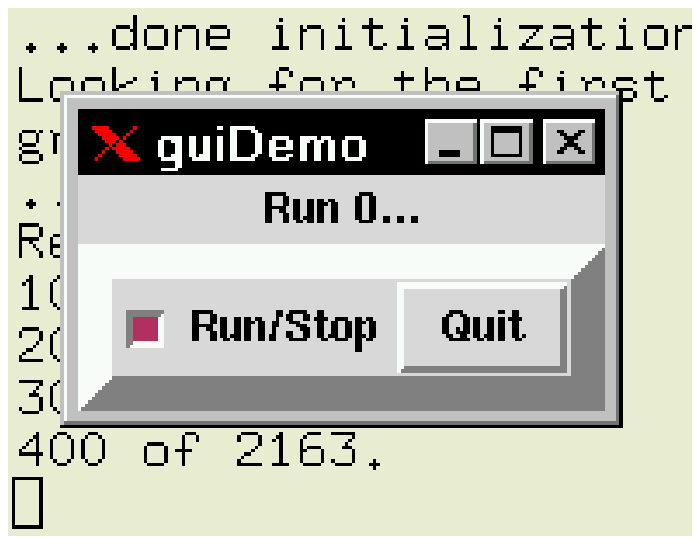}}
\end{minipage}
\begin{minipage}[b]{.32\linewidth}
c) \centerline{\epsfxsize=35mm \epsfbox{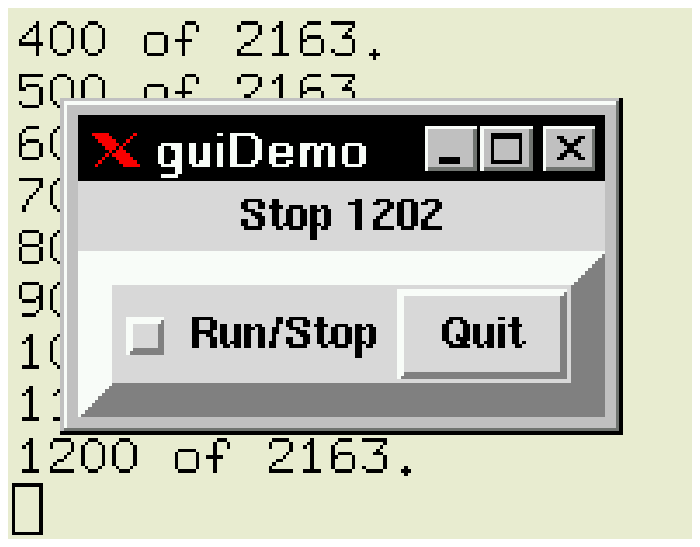}}
\end{minipage}
\caption{
\label{guiDemoSs}
   Simple demo of GUI for DIANA. During initialization, a TM program starts the
   Tcl/Tk script ``guiDemo''. The dialog appears asking the user to start
   executions, or to quit the program (a). The user starts execution (b),
   and after a short period stops execution. The dialog is waiting for the further
   user reaction (c).
}
\end{figure}

This small ``widget'' with two buttons, ``Run/Stop'' and ``Quit''
represents a simple dialog (see
Fig.~\ref{guiDemoSs}). Every
time the user clicks the ``Run/Stop'' checkbutton, the script types to the
standard output 0 or 1 (supposed to be swallowed by DIANA), 
depending on the state of the button, and reads the
window header from the standard input. If the user clicks the ``Quit''
button, then the script outputs the line \verb|Quit| to the standard output,
and exits.

The script is started by means of the operator \verb|\pipe(./guiDemo)| during
initialization of a TM program:
\begin{verbatim}
\Begin(initialization)
   . . .
   \export(guiDemo,\pipe(./guiDemo))
   \topipe(\import(guiDemo),Start/Quit?\eol())
   \if"\frompipe(\import(guiDemo))"ne"1"then
      \exit(-1)
   \endif
   \topipe(\import(guiDemo),Run\( )0...\eol())
\End(initialization)
\end{verbatim}

At this point the window appears with the title ``Start/Quit?''
(Fig.~\ref{guiDemoSs} a)) , and the execution of the TM program will be
suspended since it is blocked while waiting for the data from the pipe.

If the user clicks on the ``Quit'' button, the TM program will be finished.
If the user switches on the ``Run/Stop'' button, the TM program starts to run
(Fig.~\ref{guiDemoSs} b)).

Each time the user clicks one of the buttons (or closes the window), the script
emits some text, which becomes available for \verb|\frompipe()|.

In the beginning of the ``\verb|regular|'' section there is the following TM
code:
\begin{verbatim}
\section(regular)

\if"\ispipeready(\import(guiDemo))"eq"ok"then
\beginlabels
\label(_start)
\goto(_\frompipe(\import(guiDemo)))
\label(_0)
\topipe(\import(guiDemo),Stop\( )\currentdiagramnumber()\eol())
\goto(_start)
\label(_)
\label(_Quit)
\exit(-1)
\label(_1)
\topipe(\import(guiDemo),Run\( )\currentdiagramnumber()...\eol())
\endlabels
\endif
\end{verbatim}
If the user  switches off the ``Run/Stop'' button, this code suspends the
execution of  the TM program (Fig.~\ref{guiDemoSs} c)). The number of the
current diagram will be typed into the title.

The user can continue (checking the ``Run/Stop'' button), or interrupt
the TM program by means of the ``Quit'' button.

\section{Running external commands asynchronously}
\label{theexec}

\subsection{Parallelization, local and remote jobs and
  synchronization}
\label{parandsync}
The operator \verb|\system()| executes commands synchronously, i.e. it
waits for them to be completed and returns an exit code. In contrast,
the operator \verb|\_exec()| executes commands in the
background\footnote{On UNIX, the background jobs are programs which are
executing in such a way that they return the shell prompt while they
continue to operate. Here we use the term ``background'' in a more 
general sense.}. Similar
to the operator \verb|\pipe()|, it does
not wait for the command to be completed. It returns an empty string on success,
or some diagnostics on failure.
This operator may be used to {\bf parallelyze} the
evaluation of a process by running more than one job simultaneously.

All jobs started by the operator \verb|\_exec()| are running independently
of the TM program invoking them. To synchronize the TM program with all these
jobs, there is an operator \verb|\_waitall()|. It suspends  execution of the TM
program until timeout is expired, or until all jobs started by the
\verb|\_exec()| operator are completed (for further options see 
Sect.~\ref{relatedPar}). It returns the number of jobs which
are not yet finished, or an empty string, if there are no jobs anymore. The
single argument of the operator \verb|\_waitall()| is a timeout in millise\-conds. E.g.,
\verb|\_waitall(1000)| will return in one second the number of remaining
jobs. If the user wants to know how many jobs are not yet finished, he may
use the operator \verb|\_waitall(0)|.

To avoid overloading of
a processor, by default at each time only one job is actually running while
all the rest is waiting in the queue. The operator \verb|\_exec()| just
queues jobs.

On SMP\footnote{Symmetric MultiProcessing} computers, the optimal number of
simultaneously running jobs is the number of processors. This number can be
changed by means of the option \verb|-smp|. Thus, \verb|-smp 8|
tells DIANA to run on the computer 8 jobs simultaneously while all the
rest is queued.

In case of a cluster of computers with a current directory shared by means of
NFS e.g., the operator \verb|\_exec()| can use
DIANA servers running on other computers. To run DIANA as a server, the user
may use one of the following two options:
the option \verb|-s| tells DIANA to run a server listening to some
port, and the option \verb|-d| forces DIANA to fork out a ``daemon''
running in the background and listening to the port instead of DIANA. 
Thus, each computer on
which the user has executed the command (possibly automized, see
macro \verb|STARTSERVERS| below)
\begin{verbatim}
diana -d 1 -q
\end{verbatim}
becomes available to perform the commands queued by the operator
\verb|\_exec()| (we assume that the current directory is shared by NFS among
all computers in the cluster). Here \verb|-d 1| tells DIANA to run a daemon
accepting only one connection, and the option \verb|-q| terminates the father
DIANA process. For SMP computers the optimal argument for the \verb|-d| (or
\verb|-s|) option is the number of processors.

Running jobs are controlled by means of ``handlers''. These can be
understood as communication channels between running jobs and
DIANA. Their number is the highest possible number of simultaneously
running jobs.
By default, the operator \verb|\_exec()| will give  priority to local
handlers, i.e. if the two possibilities occur simultaneously to run
a job locally or on a remote server, the job will be started on the local
computer. 
When DIANA starts a new job, it tries to use one of the free handlers with
minimal {\em nice}. The nice is an integer number which is set for each
handler. By default, the local handler has  nice 0, while the remote
handler has  nice 1.

The nice of local handlers can be set by an optional parameter for the
option \verb|-smp|, separated by a comma from the mandatory parameter.
Thus, \verb|-smp 8|
tells DIANA to run on the computer 8 jobs simultaneously with nice 0
while \verb|-smp 8,2| tells DIANA to run these 8 jobs with nice 2.

The same with the remote server: \verb|-d 2| (or \verb|-s 2|) tells DIANA
to run a daemon (server) accepting two connections with nice 1, and
\verb|-d 2,0| (or \verb|-s 2,0|) tells DIANA
to run a daemon (server) accepting two connections with nice 0.

\subsection{Job attributes and exact syntax of operators}
\label{attribs}

Two operators \verb|\_exec()| and \verb|_waitall()| permit the user to
organize a simple parallel session for evaluations on SMP computers
and/or clusters of independent computers with shared disk space. But very
often this is not enough (see Introduction).
Therefore, each job started by the operator \verb|\_exec()| must have further {\em
attributes} (for a complete list see the end of the present section).
The first one is the ``sync'' attribute; if it is set, the
job will be started to run only after all previously started jobs are completed.
The second attribute is the ``sticky'' attribute. This attribute requires a
parameter, which gives reference to the ``master'' job to which the
``slave'' (getting this parameter) has to stick. If this attribute is set,
the ``slave'' will be performed on the same node as the ``master'' job
was running.

If the ``master'' job  fails {\em before} the ``sticky'' ``slave'' job was
started, the latter will fail too if a third attribute is set;
otherwise it will be performed independently of the result of the ``master''
job. Let's call this attribute ``stickyfail''.

There are some other attributes, which will be explained further down. The total
number of attributes is 32, but at present many of them are merely
reserved for future developments.

The operator \verb|\_execattr(attr,param)| is used to change default job
attributes. \verb|attr| is a string of length 32 in general. The position of each
character corresponds to one of the attributes. The first character
corresponds to the ``sync'' attribute, the second one  to the ``sticky'' attribute,
etc. If the corresponding character is 0, then the attribute is cleared, if
1, the attribute is set; any other character means that the attribute will
not be changed. If the string \verb|attr| is longer than 32, all extra
characters will be ignored. If it is shorter, all remaining attributes
will not be changed. Examples:
\begin{verbatim}
\_execattr(00000000000000000000000000000000,)
\end{verbatim}
clears all attributes.
\begin{verbatim}
\_execattr(1,)
\end{verbatim}
sets the ``sync'' attribute.

The second argument, \verb|param|, is used to attach a parameter to
attributes which require it. Parameters are attached consecutively 
only to those attributes
which require them and only to attributes which are set. If more than one
parameter is specified, they have to be separated by the end-of-file symbol
(EOF), which
can be obtained by the operator \verb|\eof()|.
If less parameters than set attributes are specified, they
will be cycled over all set attributes requiring a parameter.

Each job started by \verb|\_exec()| has a unique name, provided by the
user
(see below)
or defined automatically. 
The parameter for sticky attributes is the name of the ``master''
job. It is not allowed to stick to jobs which are not yet queued.
The name of the last queued job can be obtained by the
operator \verb|\lastjobname()|. So, the operator
\begin{verbatim}
\_execattr(!10,\lastjobname())
\end{verbatim}
does not change the ``sync'' attribute, raises the ``sticky'' attribute to
stick the new job to the job queued just before, and clears the ``stickyfail''
attribute.

Another example:
\begin{verbatim}
\_execattr(111,\lastjobname())
\end{verbatim}
The newly created job will be executed firstly after the ``master'' job {\em and all
previously queued jobs} are finished, secondly, on the same node\footnote{DIANA
assumes the ``node'' to be the IP address of a server, so the conception of
``nodes'' is actually supported only for clusters. For SMP, the whole computer
is assumed to be a single node.} on which the
``master'' job was performed, and thirdly
only if the ``master'' job was successful.

If the parameter for the ``sticky'' attribute is an empty string, the
last job name will be used and the above example is equivalent to
\begin{verbatim}
\_execattr(111,).
\end{verbatim}

The operator \verb|\_exec()| queues  jobs defined by  commands with arbitrary
number of command line arguments. The command and the arguments must be pushed
into a stack in  natural order by means of the operator
\verb|\push()|. 
E.g., in order to execute a command ``\verb|ls -l|'', the user has to prepare
the stack by means of the following set of operators:
\begin{verbatim}
\push(\eof())
\push(ls)
\push(-l)
\end{verbatim}
and then invokes the operator \verb|\_exec()|. 
The command line arguments appear in the stack in reverse order,
reading from top to bottom, i.e. in the stack the EOF is situated on
the bottom.  The \verb|\_exec()| operator reverses the command line
arguments from the stack such that they appear in natural order again.
The exact syntax of the \verb|\_exec()|
operator is as follows:
\begin{verbatim}
\_exec(name, attr, param).
\end{verbatim}
Here \verb|name| is the unique job name.
If it is left empty, which is recommended in general, it will be
assigned automatically.

The second and third argument have the same meaning as described above
in connection with the \verb|\_execattr()| operator. Here they are
considered as local arguments and the default ones are not
overwritten by the \verb|\_exec()|.

Besides these most important attributes we describe two more useful
ones. They are ``successcondition'' and ``restart''.

If the job is sent to a server, DIANA waits for an answer from it.
The server reports success/failure of starting the job, and after the
job is completed, the server reports the status of the job returned by the
system.

By default, the job is assumed to be successful if the server reports the
status. However, sometimes it is useful to investigate the status returned by the
system, or assume that the job is successful if it is  started by the
server (see the discussion in the Introduction).

The ``successcondition'' attribute (the fourth one)
can  accept a 
parameter which can take the following values:
\begin{itemize}
\item -2 -- the job is successful if it is started by the server;
\item -1 -- the same like without the attribute. The job is
successful if the server reports the status returned by the system.
\item from 0 to 255 -- the job is successful if the external program initiated by
the job has been
completed and returned the exit code less or equal to the specified number.
\end{itemize}

Another problem with success/failure conditions is the 
question how many times
the job should be restarted when it fails.
By default, the job is not restarted at all on failure. To change this
behaviour, a fifth attribute can be set. We  call it the ``restart''
attribute. The corresponding parameter is the number of restarts, which must be
an unsigned integer from 0 to 255.

Here is the synopsis of all attributes actually used by DIANA for the
time being:\\
\begin{center}
\begin{tabular}[t]{ccc}
{\bf No}      & {\bf Attribute}    & {\bf Parameter} \\
     1        &   sync             & --              \\
     2        &   sticky           & job name (text) \\
     3        &   stickyfail       & --              \\
     4        &   successcondition & -2{\ldots}255   \\
     5        &   restart          & 0{\ldots}255
\end{tabular}
\end{center}

\subsection{Related operators}
\label{relatedPar}
As described in Sect. \ref{parandsync}, 
the operator \verb|\_waitall()| returns control if there are no more
running jobs or if the timeout is expired. As third possibility
the operator will return control if some data are available on
a definite pipe. This pipe must be set by
 means of the operator
\verb|\setpipeforwait(rpipe)|, where \verb|rpipe| is the PID returned by the
operator \verb|\pipe()|, or 0. In the latter case the operator
\verb|\_waitall()| will return  control if there is something typed on
the keyboard.

If control is returned due to available data from the pipe,
the number returned by the operator \verb|\_waitall()| is preceded by a
minus sign. If all jobs finish simultaneously with
data occurring in the pipe  only a minus sign is returned.

The operator \verb|\getpipeforwait()| returns the PID of the pipe which has been set by
\verb|\setpipeforwait(rpipe)|
or an empty string if there is no active pipe for \verb|\_waitall()|.

\verb|\pingServer(ip)| checks if a DIANA server is available on a given IP address.
The argument \verb|ip| is the IP address in standard dot notation. The
operator returns
\verb|alive| if the server responds, or an empty string if the server
cannot be found or does not respond.

\verb|\killServer(ip)| kills the DIANA server running on the specified IP address.
The argument \verb|ip| is again the IP address in standard dot
notation. The operator returns
\verb|ok| on success, or an empty string on failure.

\verb|\killServers()|  kills all running DIANA servers.
It returns an empty string on success, or \verb|none| on failure.

\verb|\lastjobname()| returns the name of the last job started by the
operator \verb|\_exec()|.

\verb|\lastjobnumber()| returns the number of the last job started by
the operator \verb|\_exec()|.

\verb|\whichIP(jobname)| returns the IP address as 8 hexadecimal digits or an empty
string on failure.

\verb|\whichPID(jobname)| returns the PID as a hexadecimal number or an empty
string on failure.

\verb|\getnametostick(jobname)| returns the name of the ``master'' job to
which the specified ``sticky'' job \verb|jobname| has to stick. On
failure it
returns an empty string.

\verb|\jobtime(jobname)| returns 8 hexadecimal
digits, the job running time in millise\-conds. If the job is not yet running,
the operator returns ``00000000''.

\verb|\jobstime(jobname)| returns 8 hexadecimal digits, the job starting time
in millise\-conds. The starting time is counted from the moment of the DIANA job
queuing mechanism being activated till the time the job starts to
run. 
If the job is not yet started, it returns
``00000000''.

\verb|\jobftime(jobname)| returns 8 hexadecimal
digits, the job finishing time
in millise\-conds. The finishing time is counted from the moment of the DIANA job
queuing mechanism being activated till the job is completed.  
If the job is not yet finished, it returns
``00000000''.

\verb|\rmjob(jobname)| removes the named job. If the job is running, the
related external programs will automatically be killed by the signal SIGKILL.
The operator returns an empty string independently of the result.

\verb|\clearjobs()| clears all jobs. It returns an empty string on success,
or the string ``Client is dead.''

\verb|\jobstatus(jobname)| returns the status of a specified job.
The status is given in terms of 8 hexadecimal digits. The first two
digits 
stand for
the \verb|exit()| argument of the external program if the program
exited normally, otherwise they are ``00''.
The third and fourth digits are the following:
``00'' -- normal exit, ``01'' -- exit because of non-caught signal, ``02'' -- the
job was not run, ``03'' -- no such job, ``04'' -- job is finished, but its
status is lost (often occurs after the debugger being detached); otherwise -- ``ff''.    
If the job was removed by the operator \verb|\rmjob()|, the third digit will
not be ``0'' but ``1''.

The last 4 digits stand for the number of the signal  that
caused  the external program to terminate if the fourth digit
was ``1'', otherwise ``0000''.

\verb|\startjobtimeout(timeout)| sets the timeout for attempts 
to start jobs (mil\-li\-se\-conds).
After the timeout is expired and the job is not yet running,
it will be moved to another handler if
the total number of the timeout hits is less then 10.
      If the number of  timeout hits is 10, the job fails.
      Time is counted  from the moment of
        the job being sent to the remote (or local) server.
      The operator returns the 
previous value of the timeout or an empty string on failure.

\verb|\gconnecttimeout()|
sets the timeout for the connection with a remote server (millise\-conds).
It returns the previous value of the timeout, or an empty string on failure.

\verb|\jobhits(jobname)| returns 6 hexadecimal digits.
The first two digits stand for the number of times
the job was restarted due to the start timeout expiration.
The third and fourth digits stand for the number of  times
the job was restarted after failures
and the last two digits represent the
current placement of the job:
``01'' -- the job is in the main queue,
``02'' -- the job is in the synchronous  queue,
``03'' -- the job is finished,
``04'' -- the job failed,
``05'' -- the job runs,
``06''  -- the job is ready to run\footnote{I.e. all conditions to start
the job are fulfilled and the job will be started as soon as some proper
handler becomes available.} as a non-sticky job,
``07''  -- the job is ready to run as a sticky job.

\verb|\failedN()| returns number of failed jobs, or an empty string, on failure.

\verb|\hex2sig(sig)| converts DIANA's internal signals' hexadecimal representation
to local system ones (decimal numbers).

\verb|\sendsig(jobname,signal)| sends a signal to the external program
which was initiated by the specified job. Note, the signal will be sent to {\em
all} external programs initiated by the job\footnote{The external
program initiated by the job runs in a new session (UNIX - specific term). Actually, the signal
will be sent to the {\em process group} corresponding to the initial external
program.}.
It expects that the  \verb|signal| is either a
symbolic name SIG{\ldots}  (the SIG prefix may be omitted), or a number.
  All possible symbolic names are (a subset from the conventional UNIX
  signal identifiers):
  SIGHUP SIGINT SIGQUIT SIGILL SIGABRT SIGFPE SIGKILL SIGSEGV SIGPIPE
  SIGALRM SIGTERM SIGUSR1 SIGUSR2 SIGCHLD SIGCONT SIGSTOP SIGTSTP
  SIGTTIN SIGTTOU SIGBUS SIGPOLL SIGPROF SIGSYS SIGTRAP SIGURG SIGVTALRM
  SIGXCPU SIGXFSZ SIGIOT SIGEMT SIGSTKFLT SIGIO SIGCLD SIGPWR SIGINFO
  SIGLOST SIGWINCH.
If the signal is specified by a number, the signal will be sent as it is. If
the signal is specified by a symbolic name, DIANA will convert it into an
internal representation transferred to the remote server. This may be
useful if the numerical representation of signals on the remote system
differs from the local one.
 The operator returns an empty string on success; on failure it
returns:
``10'' -- job is not run, ``20'' -- can't read a signal from
DIANA; less then 10:
          ``01'' -- an invalid signal was specified,
          ``02'' -- the pid or process group does not exist,
          ``03'' -- the  process  does  not have permission
          to send the signal,
          ``04'' -- unknown error.

\verb|\getJobInfo(jobn, format)| returns information about the job number
\verb|jobn|, formatted according to \verb|format|. The latter is a string of
type \verb|n1/w1:...:nn/wn'|, where \verb|n#| is a field number and \verb|w#|
is its width.  If \verb|w#|$\,\,>0$, the text in the field will be right
justified and padded on the left with blanks. If \verb|w#|$\,\,<0$, then the text
in the field will be left justified and padded on the right with blanks.
If \verb|w#|=0, then the width of the field will be of the text length.
Here is a complete list of all available fields:
\begin{enumerate}
\item current placement, see above;
\item return code, 8 hexadecimal digits, see above;
\item IP address (in dot notation);
\item PID (decimal number);
\item attributes (decimal integer);
\item timeout hits (decimal number), see above;
\item fail hits (decimal number), see above;
\item max fail hits (decimal number), see above;
\item errorlevel (decimal number), see above;
\item the job name;
\item number of command line arguments;
\item all command line arguments starting from a second one (spaces separated);
\item all command line arguments starting from a third one (spaces separated);
\item all command line arguments starting from a fourth one (spaces separated).
\end{enumerate}

If \verb|n#| is zero or negative its absolute value is interpreted as
the position of a single
command line argument. 

Example:
\verb|\getJobInfo(1, 4/10:3/15)|
returns the PID of the external program of the first job and the 
IP address of the
computer on which the job runs. PID will be returned as a decimal number
padded by spaces on the left to fit a field of width 10, and IP will be
returned in dot notation padded by spaces on the left to fit a field of
width 15.

\verb|\tracejobson(filename)| starts jobs tracing 
(for details see \\
http://www.physik.uni-bielefeld.de/\verb|~|tentukov/parallel\_demo.tar.gz)
into files \linebreak
(``filename.var'' and
``filename.nam''),
   returns an empty string. 

\verb|\tracejobsoff()| ceases tracing. It returns an empty string.
After it has stopped, tracing cannot be continued until all
jobs are cleared by
means of the operator \linebreak \verb|\clearjobs()|.

\subsection{Example of high-level usage of parallel processing facilities}
\label{parascript}

Very often the external program may  run in a subshell as a conventional
UNIX command. For such cases the following high-level TM function
\verb|\exec(command)| can be defined:
\begin{verbatim}
\function exec cmd;
 \push(\eof())
 \push(sh)
 \push(-c)
 \push(\get(cmd))
 \return(\_exec(,00011,255\eof()4))
\end
\end{verbatim}
The function starts the job consisting of some (maybe composed) standard
UNIX command. The job will be successful if the command returns some
exit code (under  UNIX there is no possibility to have an exit
code  higher then 255). On failure the job will be restarted up to four
times (the number 4 after \verb|\eof()|).

The function \verb|\stick(command)|:
\begin{verbatim}
\function stick cmd;
 \push(\eof())
 \push(sh)
 \push(-c)
 \push(\get(cmd))
 \return(\_exec(,11110,\eof()-2))
\end
\end{verbatim}
performs the command \verb|cmd| only after all earlier jobs are
completed (the first 1 of the second argument of \verb|\exec()|).
This function in general is used to collect all produced files.
That is why it performs \verb|cmd| on the same computer as the
preceding job (the second 1; since the job name to stick to is an empty
string, the last job name will be used).
The job started by this function is successful if it is started by the
server (the fourth 1, and the corresponding parameter is ``-2''). The
job will not be restarted on failure (the last 0).

The function \verb|\wait(timeout)|:
\begin{verbatim}
\function wait to;
 \while"\let(r,\_waitall(\get(to)))"ne""do
  \message(\get(r)\( jobs are not finished))
 \loop
\end
\end{verbatim}
suspends execution of the TM program
until all jobs are completed. Each ``\verb|timeout|'' millise\-conds the function
reports the number of jobs which are not yet finished.

These three functions can be used to organize a simple parallel session with
proper synchronization.
As an example, let us consider the following script \verb|runf|, which
is used to run FORM jobs in parallel after the FORM input has been
produced in a folder (see \cite{FORM}, p.~158):
{\footnotesize
%\begin{figure*}[t]
\begin{verbatim}
#!diana -smp 1 -c runpar.tml
\STARTSERVERS(phya25,phya26,phya27,phya28)
\system(echo > log.all)

\REPEAT(N)
\exec(form -d i=\get(N) do.frm > /tmp/log.\get(N))
\stick(cat /tmp/log.\get(N) >> log.all)
\stick(rm /tmp/log.\get(N))
\ENDREPEAT()

\wait(2000)
\end{verbatim}
%\caption{\label{thescript}}
%\end{figure*}
}
\noindent
The user enters: \verb|runf 186 200 |, and the system executes:\\
\verb|diana -smp 1 -c runpar.tml runf 186 200|.

The file \verb|runpar.tml| (see Appendix \ref{runpar})
contains definitions of various TM functions and
some settings; in particular it redefines the comment character as
\verb|#|.
The following is the structure of the file \verb|runpar.tml|:
\begin{verbatim}
messages disable
esc character = \
comment character = #
output file =""
debug off
only interpret
\begin translate
\end{verbatim}

\begin{minipage}{0.5\linewidth}
 \em  Definitions various functions and macros, \nopagebreak in \nopagebreak
particular \nopagebreak
 \verb|\exec()|,\nopagebreak
 ~\verb|\wait()|,\\
 \verb|\REPEAT()|,
 \verb|\ENDREPEAT()|
 and
\verb|\STARTSERVERS()|
\end{minipage}

\begin{verbatim}
\program
\-
\SET(_CMDLN)(\CMDLINE(1))
\RMARG(1)
\{\include(\GET(_CMDLN))\-
\setout(null)
\}
\end translate
\end{verbatim}

The main program switches off the output (by means of the directive \verb|\-|),
switches off ``hash'' regime for arguments by means of the directive
``\verb|\{|'' (see below), stores the first command line argument\linebreak[4]
(``\verb|\SET(_CMDLN)(\CMDLINE(1))|''), removes it  (``\verb|\RMARG(1)|'')
and includes the file the  name of which is the first command
line argument \\ (``\verb|\include(\GET(_CMDLN))|'').

By default, DIANA provides for the function and operator arguments some
special ``hash'' regime: all spaces and ends of line are ignored.
The directive ``\verb|\{|'' cancels this regime, the directive ``\verb|\}|''
activates it again. Without this directive all commands with spaces must be
quoted by the quotation operator ``\verb|\()|'' which is not convenient for a script.

The macro \verb|\STARTSERVERS(list)| checks if each server from \verb|list|
is working
and if not starts the server by means of the \verb|ssh|. For example, for
the host phya26 the following command will be performed:
\begin{verbatim}
ssh phya26 cd CD ; diana -d 1 -q
\end{verbatim}
where ``\verb|CD|'' is a current directory, e.g. \verb|/home/user/jobs|.

The operator \verb|\system(echo > log.all)|
is used to produce an empty file \linebreak ``\verb|log.all|''.

All the instructions between \verb|\REPEAT(N)| \verb|...| \verb|\ENDREPEAT()| are cycled
with \linebreak \verb|N=186,...,200|.
We assume that there is some folder file, say, \verb|tt.in| with FORM input
for each diagram. The FORM program \verb|do.frm| evaluates a
diagram by virtue of including a fold from the folder \verb|tt.in| via
an instruction like
\verb|#include tt.in # n'i'|. The macro definition \verb|i| comes from
the command line \verb|form -d i=\get(N) do.frm|, where \verb|\get(N)|
runs from 186 to 200. Each FORM job saves the result to the local
directory, but the corresponding concatenation is performed by \verb|\stick(cat ...)|
on the same computer. At the end, all results will be collected in the file
\verb|log.all|, and all intermediate files \verb|\tmp\log.#| will be removed.

After all jobs are queued, the function \verb|\waitall(2000)| will report
every 2 seconds how many jobs are not yet completed.

This example is available at\\
http://www.physik.uni-bielefeld.de/\verb|~|tentukov/parallel\_demo.tar.gz.

In the archive there are two directories: demo1 and demo2. The example
is situated in demo1; see the Readme file for details of
installation. For a more realistic (and complicated)
example containing jobs controlling and monitoring facilities see  demo2.

\section*{Acknowlegement}
M.~Tentukov is grateful for financial suuport by the DFG under project
no. FL~241/4-2

\appendix
\section*{Appendix}

\section{The listing of the Tcl/Tk script guiDemo}
\label{guiDemo}
The following Tcl/Tk script is supposed to be placed into the
executable file \verb|guiDemo|:
\begin{verbatim}
#! /bin/sh
#\
exec wish "$0" "$@" -geometry +200+300

# catches destroy window event:
wm protocol .WM_DELETE_WINDOW {puts "Quit";exit 0}

label .title
pack .title

# create relief canvas:
frame .buttons -bd 10 -relief raised

#Button "Run/Stop":
global checkb
frame .buttons.run
checkbutton .buttons.run.run -text "Run/Stop"\
  -command {puts "$checkb"; flush stdout;\
            .title configure -text [gets stdin]}\
             -variable checkb
pack .buttons.run.run
pack .buttons.run -side left

#Button "Quit"
frame .buttons.quit
button .buttons.quit.quit -text "Quit" -command\
  {puts "Quit";exit 0}
pack .buttons.quit.quit
pack .buttons.quit -side left
pack .buttons

# Read initial header:
.title configure -text [gets stdin]
#End:
\end{verbatim}

The script creates a window with two buttons. The check button ``Run/Stop''
has two states, ``selected'' and ``deselected''. The global variable
\verb|checkb| is set to indicate whether or not this button is selected.
Every time the user clicks the ``Run/Stop'' checkbutton, the script types to the
standard output the value of \verb|checkb| and reads the
window header from the standard input.
The ``Quit'' button, when pressed, outputs the line ``\verb|Quit|'' to the
standard output and exits from the script. The same occurs when the user
closes the window.

\section{The listing of the file ``runpar.tml''}
\label{runpar}
The following listing is typed in {\tt typewriter} while comments are
typed using {\it italic}.

{\footnotesize
{
\catcode`\|=0
|catcode`|\=11
|begin{alltt}
{|it{}Preamble:}
messages disable
esc character = \
comment character = #
output file =""
debug off
only interpret
\begin translate

{|it{}The function returns {|bf true} if 
the argument is a number:}
\function isnumber str;
  \if"\get(str)"eq""then
     \let(res,)
  \else
     \let(sk,\getcheck())
     \setcheck(0123456789)
     \let(res,\check(\get(str)))
     \setcheck(\get(sk))
  \endif
  \return(\get(res))
\end

\function wait to;
 \while"\let(r,\_waitall(\get(to)))"ne""do
  \message(\get(r)\( jobs are not finished))
 \loop
\end

{|it{}Begin macrodefinition of the
REPEAT environment}
\DEF(REPEAT)
\IFSET(__REP)
\ERROR(Nested REPEAT!)
\ENDIF
\SET(__REP)(1)

{|it{}Checking command line arguments:}
\if"\exist(__From)"eq"false"then
   \let(__From,\cmdline(1))
\endif
\if"\exist(__To)"eq"false"then
   \let(__To,\cmdline(2))
\endif
\beginlabels
\label(again)
\while"\isnumber(\get(__From))"ne"true"do
   \let(__From,\read('\get(__From)'
      \( is not an integer, enter new FROM:)))
\loop

\while"\isnumber(\get(__To))"ne"true"do
   \let(__To,\read('\get(__To)'
      \( is not an integer, enter new TO:)))
\loop

\if"\numcmp(\get(__From),\get(__To))"eq">"then
   \message(\get(__From)>\get(__To)!)
   \let(__From,)\let(__To,)
   \goto(again)
\endif
\endlabels

{|it Pre-processor SCAN is used since macro argument }\#(1)
{|it cannot be used in }\if{|it operator:}
\SCAN(\if"\#(1)"eq""then)
   \let(_iND,_index){|it Default iterator}
\else
   \let(_iND,\#(1)){|it Use argument as iterator}
\endif

\let(\get(_iND),\get(__From))
\do {|it Start the main loop}
\ENDDEF
{|it{}End macrodefinition of the
REPEAT environment}

{|it{}Macrodefinition of the ENDREPEAT
environment, rather simple:}
\DEF(ENDREPEAT)
{|it{}End of the main loop}
\while"\numcmp(\inc(\get(_iND),1),\get(__To))"ne">"loop
\UNSET(__REP)
\ENDDEF

\function exec cmd;
   \push(\eof())
   \push(sh)
   \push(-c)
   \push(\get(cmd))
  \return(\_exec(,00011,255\eof()4))
\end

\function stick cmd;
   \push(\eof())
   \push(sh)
   \push(-c)
   \push(\get(cmd))
   \return(\_exec(,11110,\eof()-2))
\end

{|it{}Attention! The following macro assumes that
the executable file ``diana'' is availavle
from the system path set in the PATH variable,
and the current directory is shared by NFS:}
\DEF(STARTSERVERS)
   \let(_cmd,\( 'cd )\asksystem(pwd,)\( ; diana -m d -d 1,1 -q'))
   \FOR(_srv)(\*){|it Loop on all arguments}
    {|it Remove possible spaces from the server name:}
     \let(_n,\delete(\_srv(),\( )))
     \let(_ip,)
     \if"\get(_n)"ne""then
       {|it Get IP address:}
        \let(_ip,\getip(\get(_n)))
     \endif
     \if"\get(_ip)"ne""then
       {|it Is the server alive?:}
        \if"\pingServer(\get(_ip))"eq""then
          {|it Not yet. Start it:}
           \message(\(Starting server at )\get(_ip)...)
           \system(\(ssh )\get(_ip)\get(_cmd))
        \endif
     \endif
   \ENDFOR
 {|it Servers may be started successfully, but 
   network connection establishing may take some time.
   So try to ping servers, and if it does not respond,
   wait one second:}
   \FOR(_srv)(\*)
     \let(_n,\delete(\_srv(),\( )))
     \let(_ip,)
     \if"\get(_n)"ne""then
        \let(_ip,\getip(\get(_n)))
     \endif
     \if"\get(_ip)"ne""then
        \if"\pingServer(\get(_ip))"eq""then
           \system(sleep 1)
        \endif
     \endif
   \ENDFOR
\ENDDEF

{|it{}Main program:}
\program
\-{|it Switch off output}
{|it{}Store first command line argument:}
\SET(_CMDLN)(\CMDLINE(1))
\RMARG(1){\|it Remove it}
{|it{}Now command line arguments available 
from the script are counted in correct order}
{|it{}Include a script body:}
\|{\include(\GET(_CMDLN))\-
{|it{}Switch off output and close output
possibly opened by the script:}
\setout(null)
\|}
\killServers(){|it Kill all servers}
\end translate
|end{alltt}
}
}

\end{document}